\documentstyle[preprint,aps,epsf]{revtex}
\begin{document}
%%%%%%%%%%%%%%%%%%%%%%%%%%%%%%%%%%%%%%%%%%%%%%%%%%%%%%%%%%%%%%%%%%%%%%

\draft
\title{
  Dispersion relations to oscillatory reaction-diffusion systems with
  the self-consistent flow
}
\author{
  Hiroyasu Yamada\footnote[1]{Present address:
    Research Institute for Electronic Science, Hokkaido University,
    Sapporo, 060-0812, JAPAN}
  and Toshiyuki Nakagaki
}
\address{
  Bio-Mimetic Control Research Center,
  The Institute of Physical and Chemical Research (RIKEN)
  Shimoshidami, Moriyama, Nagoya, 463-0003, JAPAN
}
%\thanks{}
%\date{\today}
\maketitle

\begin{abstract}
  Dispersion curves to a oscillatory reaction-diffusion system with
  the self-consistent flow have obtained by means of numerical
  calculations. The flow modulates the shape of dispersion curves and
  characteristics of traveling waves. The point of inflection which
  separates the dispersion curves into two branches corresponding to
  trigger and phase waves, moves according to the value of the
  advection constant. The dynamics of phase wave in
  reaction-diffusion-advection equations has been studied by limit
  cycle perturbations. The dispersion relation obtained from the phase 
  equation shows that the competition between diffusion and advection
  constants modulates the oscillation frequency from the bulk
  oscillation in the long-wave dynamics. Such a competition implies
  that phase waves with the flow have a wider variety of dynamics than 
  waves without the flow.
\end{abstract}


%---------------------------------------------------------------------
\section{Introduction}
\label{sec: Introduction}

In biological systems of cell population, taxis induces cooperative
movement of cells or the flow of cellular mass. Cells behave according 
to chemical and physiological signals given by themselves. Two
well-known pattern formations for population dynamics of cells are the 
aggregation of cellular slime molds and the branching growth of
bacterial colonies. Slime mould cells secrete signaling chemicals
responding to extracellular chemicals, and move towards increasing
some chemical concentration. They aggregate to centers of the wave
pattern of chemicals with dendritic streaming lines and eventually
form slugs\cite{Bonner94,Devreotes89,Tomchik81}.
Morphology of bacterial colonies depends on agar concentration and
nutrient level. Bacteria are swimming toward high concentration of
nutrients, and form various colonial patterns as concentric rings and
dendritic branchings\cite{Burdene,BenJacob,Matsushita}.
Mechanisms of these pattern formations have been modeled by taxis
equations or reaction-diffusion systems with the cellular flow.

For one of protozoan myxomycete, the {\it Physarum\/} plasmodium,
which is a unicellular organism, a reaction-diffusion-flow model is
presented in relation to control mechanism of amoeboid movements.
The flow of endoplasm and chemicals induced by protoplasmic streaming
in {\it Physarum\/} plays an important role in individual behavior
under a large plasmodium state. It is not difficult to look for
reaction, diffusion and flow couplings in various biological process,
more and more. The self-consistent flow as above is realized by
biological active process, and is a different character from
physico-chemical ones. Effects of the self-consistent flow on
simple reaction-diffusion systems are to be clarified in order to
study a mechanism of functional self-organization of multicellular
systems.

In the present article, we discuss basic effects of the flow on
oscillatory phenomena in a reaction-diffusion system with the
self-consistent flow. Although we dealt with a model for the
{\it Physarum\/} plasmodium, it is expected that the results obtained
here are applicable to other biological systems with the flow.
We carry out numerical calculation of the dispersion relation in the
system, and show effects of the flow on traveling plane waves.
Then we derive the dynamics of phase waves from
reaction-diffusion-advection equations in general form by means of
limit cycle perturbations. In the coefficient of the gradients of the
phase, the advection terms compete with diffusion terms.

%---------------------------------------------------------------------
\section{Model equations for the plasmodium}
\label{sec: Model equations}

The plasmodium of {\it Physarum\/} has a cytoplasmic cortex
(ectoplasm) filled with endoplasm. The ectoplasm shows contractile
oscillation everywhere within the organism
\cite{Ueda78,NaibMajani82,Ishigami86},
and the contraction causes intracellular streaming of
the endoplasm \cite{Kamiya59}.
The contraction-relaxation behavior is regulated by metabolic cycles
of chemical oscillators \cite{Grebecki78,Baranowski82}.
Metabolic chemicals transported by endoplasmic streaming, affect
chemical oscillators \cite{Yoshimoto78,Miyake91,Nakagaki96}.

A model of contractile and motile dynamics of the {\it Physarum\/}
plasmodium is represented by conservation of the cytoplasmic mass and
reaction-diffusion system of metabolic elements with endoplasmic
streaming\cite{Teplov,Yamada99}.
Under some assumptions, oscillatory dynamics in the plasmodium is
governed by the following reaction-diffusion equations with advection
terms:
\begin{equation}
  \frac{\partial {\bf u}}{\partial t}
  + M \vec{\nabla} {\bf u} \cdot \vec{\nabla} {\bf u}
  = {\bf f}({\bf u}) + D \vec{\nabla}^2 {\bf u},
  \label{eqn: rda}
\end{equation}
where ${\bf u}$ is a $N$-component vector of reaction species which
are separated into two types of metabolic elements, free chemicals and
non-free ones. While the former is transported by the protoplasmic
streaming, the latter is bound or stored at some cytoplasmic
structure. Reaction kinetics ${\bf f}$ denotes the metabolic
oscillation which emerges from the coupling of two types of reaction
elements mentioned above, and the system is assumed to have a limit
cycle orbit. The quantity $D$ is a diagonal matrix of diffusion
constants. A tensor $M$ represents advection coefficients, here $M
\vec{\nabla} {\bf u}$ is a flow vector for each reaction
component. This system has a generic form of reaction-diffusion
equations with the self-consistent flow, and we derive the phase
dynamics from eq.~\ref{eqn: rda} in section \ref{sec: Phase dynamics}.

\begin{figure}[p]
  \vfil
  \begin{center}
    \leavevmode
    \epsfxsize=0.8\textwidth
    \epsfbox{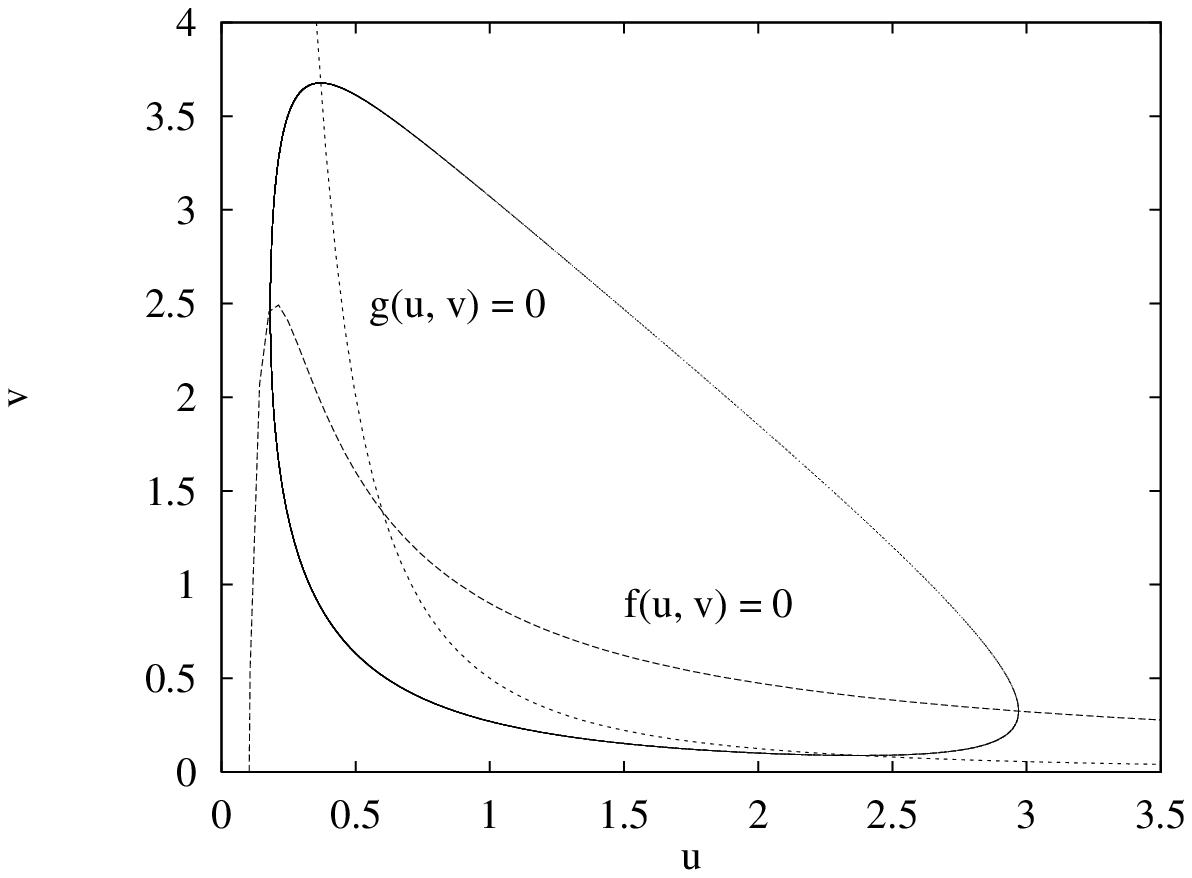}
    \caption{The limit cycle of the Schnackenberg's tri-molecular two
      species model.}
    \label{fig: limit cycle}
    % PostScript filename: orbit.ps
  \end{center}
\end{figure}

In our numerical calculations, we use a more simple two-variable
model\cite{Yamada99}:
\begin{eqnarray}
  &&\frac{\partial u}{\partial t} + \vec{w} \cdot \vec{\nabla}
  = f(u, v) + D \vec{\nabla}^2 u,
  \nonumber\\
  &&\frac{\partial v}{\partial t} = g(u, v),
  \label{eqn: 2rda}
\end{eqnarray}
where $u$ and $v$ are the concentrations of a free chemical substance
and a bound/stored one, respectively. We adopt the Schnackenberg's
tri-molecular two species model for reaction kinetics of
chemicals\cite{Schnackenberg79},
\begin{equation}
  f(u, v) = a - u + u^2 v,\quad g(u, v) = b - u^2 v,
  \label{eqn: tmts}
\end{equation}
here $a$ and $b$  are positive constants. In the spatially homogeneous
conditions ($D = 0$ and $\vec{w} = \vec{0}$), the system has a stable
limit cycle as shown in Fig.~\ref{fig: limit cycle} for
$b - a > (a + b)^3$.
The quantity $D$ is the diffusion constant of the free chemical $u$.
The velocity of the endoplasmic flow $\vec{w}$ is determined by the
concentration of the metabolic chemical as
\begin{equation}
  \vec{w} = q \vec{\nabla} u,
  \label{eqn: flow}
\end{equation}
and thus the flow is self-consistent. Here $q$ depends on the
permeability and the mechanism of intracellular pressure. In the
following calculation, $q$ is assumed to be constant.

%---------------------------------------------------------------------
\section{Dispersion relation}
\label{sec: Dispersion relation}

The numerical calculations were carried out for eq.~\ref{eqn: 2rda} on 
a ring, that is a one-dimensional region with periodic boundary
conditions. Using the limit cycle, we initiate a pulse traveling on
the ring, and solve eq.~\ref{eqn: 2rda} until the solution becomes
periodic in time. After we measured the rotating period of the
traveling pulse on the ring, we repeat the calculation for rings of
different lengths. As above, the dispersion relation has been obtained 
for periodic wave trains with the stable propagation.

In the measurements, parameters have been set as 
$a = 0.1$, $b = 0.5$, and $D = 1.0$.
We have used the explicit Euler's method for reaction terms, upwind
differencing method for advection terms, and the implicit method for
diffusion terms. The data have been recorded for waves in stationary
propagation.

\begin{figure}[p]
  \vfil
  \begin{center}
    \leavevmode
    \begin{minipage}[t]{\textwidth}
      \epsfxsize=0.8\textwidth
      \epsfbox{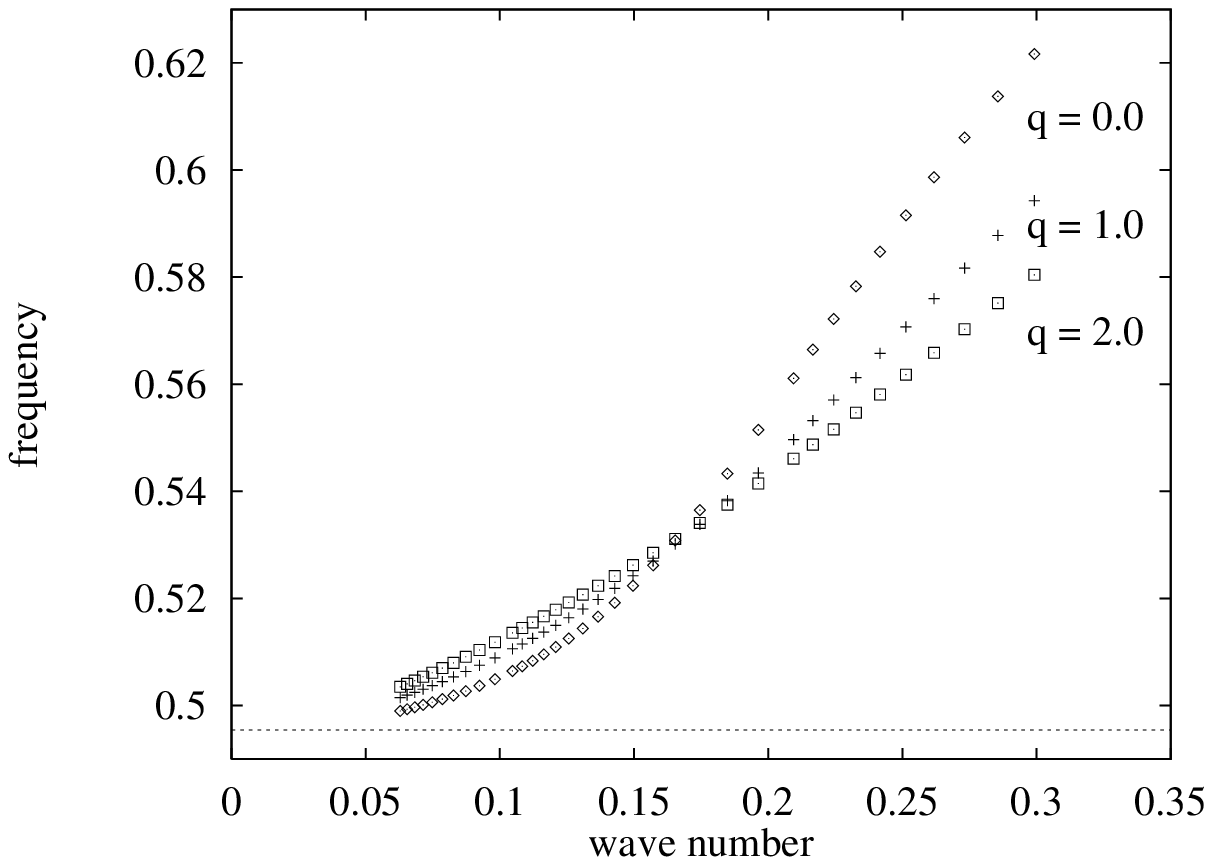}
    \end{minipage}
    \vskip 12pt
    \begin{minipage}[t]{\textwidth}
      \epsfxsize=0.8\textwidth
      \epsfbox{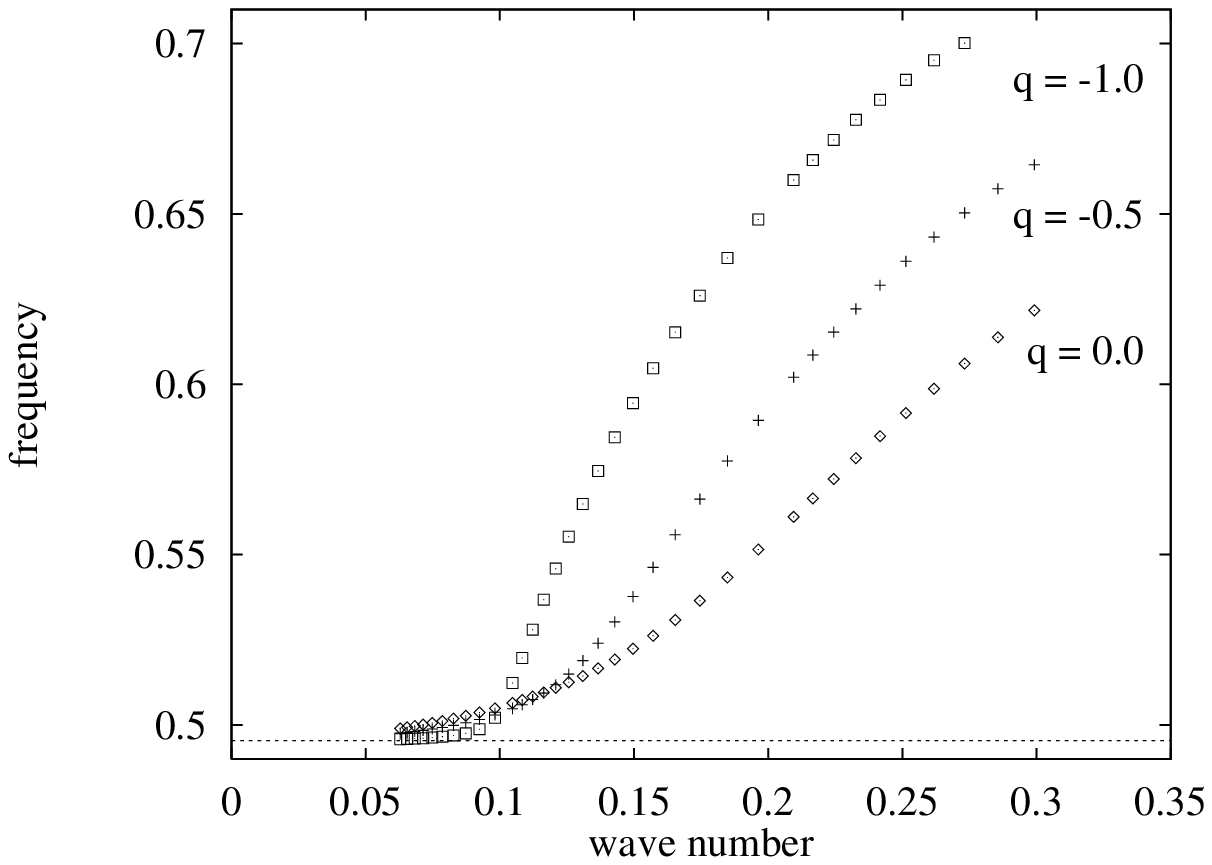}
    \end{minipage}
    \vskip 24pt
    \caption{Dispersion curves obtained by numerical calculations.
      The oscillation frequency $\omega$ monotonically increase with
      the wave number $k$, and has the bulk frequency
      $\omega_0 = 0.495$ in the limit of $k \to 0$.}
    \label{fig: k-omega}
    % PostScript filename: k-omega.ps
  \end{center}
\end{figure}

\begin{figure}[p]
  \vfil
  \begin{center}
    \leavevmode
    \begin{minipage}[t]{\textwidth}
      \epsfxsize=0.8\textwidth
      \epsfbox{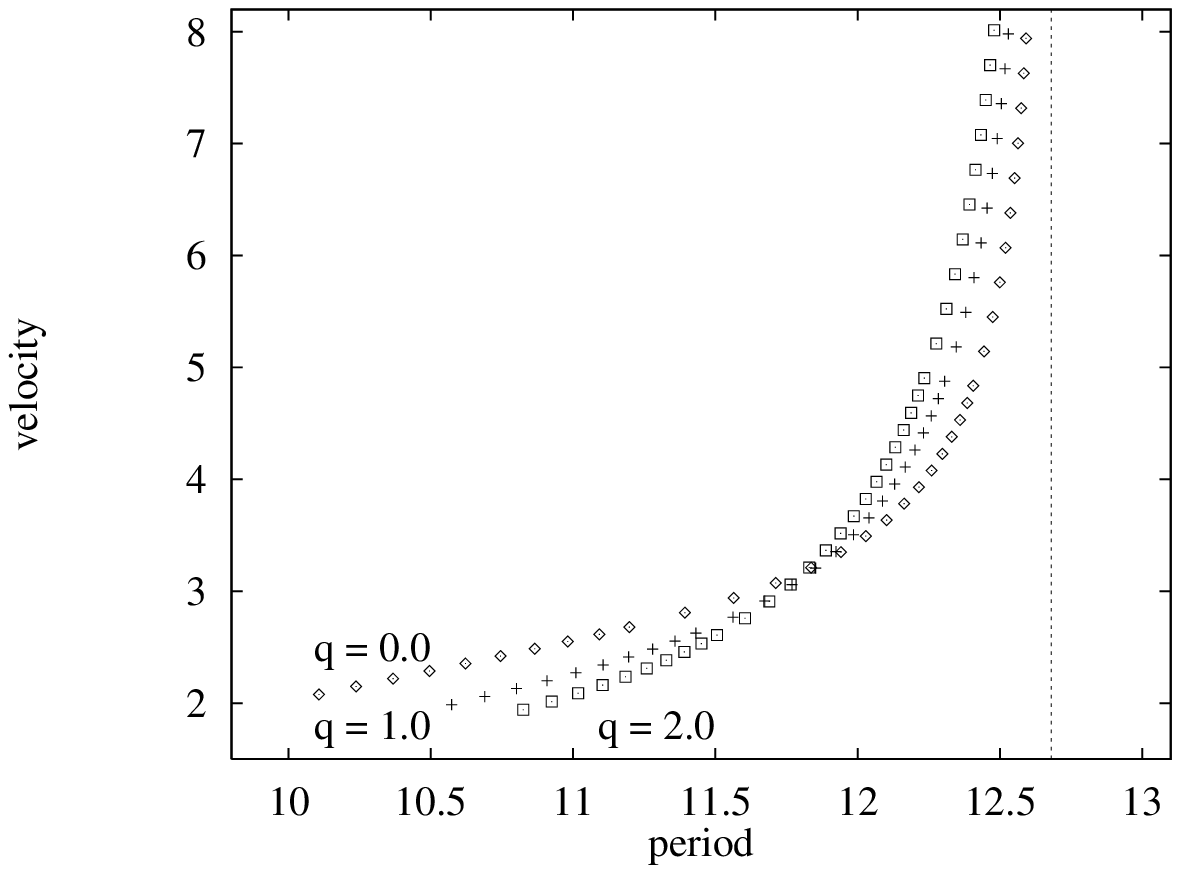}
    \end{minipage}
    \vskip 12pt
    \begin{minipage}[t]{\textwidth}
      \epsfxsize=0.8\textwidth
      \epsfbox{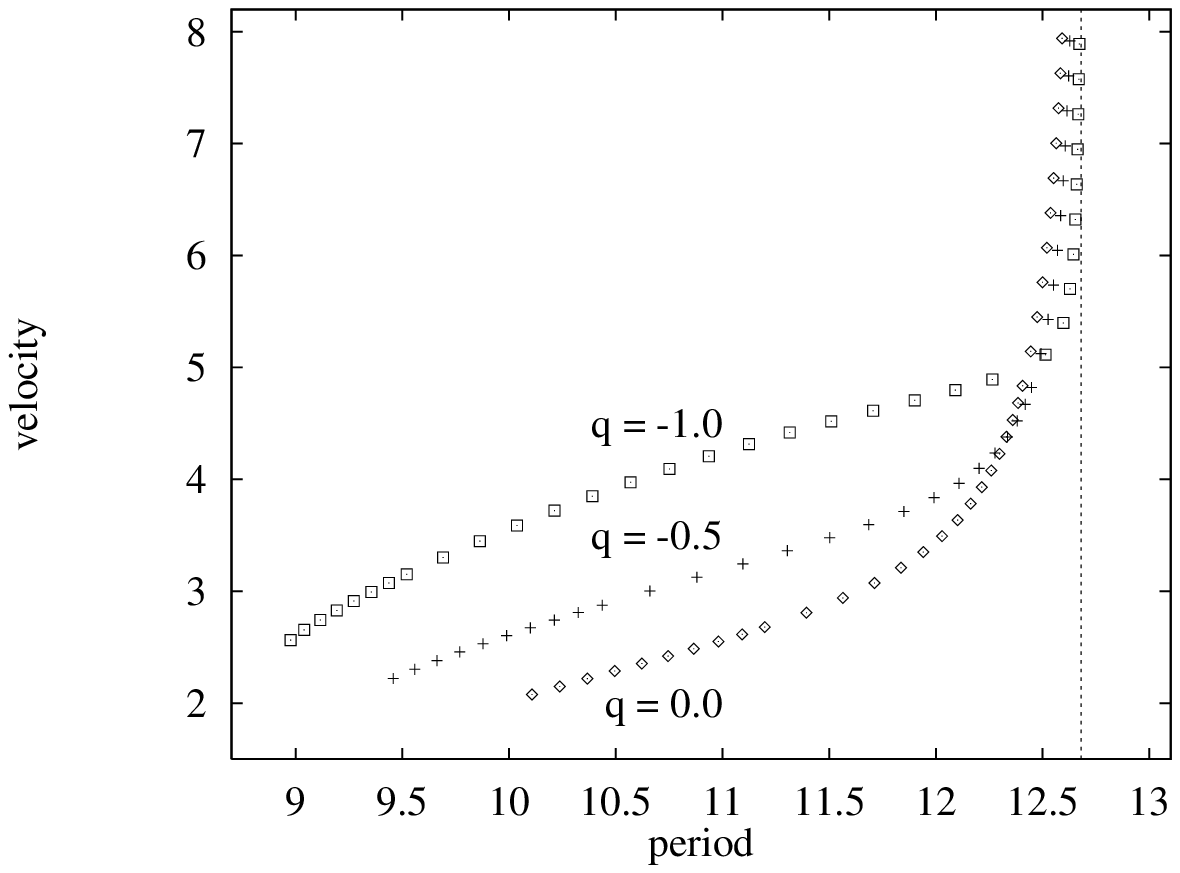}
    \end{minipage}
    \vskip 24pt
    \caption{The propagating velocity $v$ is determined as the phase
      velocity $\omega / k$. The velocity becomes large and goes to
      infinity at the bulk period $\tau_0 = 12.7$. For negative value
      of $q$, the curves have large convex regions in period
      corresponding to trigger waves.}
    \label{fig: tau-v}
    % PostScript filename: tau-v.ps
  \end{center}
\end{figure}

Dispersion curves in the reaction-diffusion-advection
system \ref{eqn: 2rda} are shown in
Fig.~\ref{fig: k-omega} (wave number vs frequency) and
Fig.~\ref{fig: tau-v} (period vs velocity)
for various values of the advection
constant $q$. Variations in the self-consistent flow make the
propagation of plane waves quite changed. The inflection point which
separates two branches corresponding to trigger waves and phase waves
\cite{Aliev94,Polezhaev95},
moves according to the flow. It is the point that the variation of
propagating behavior depends on the wave length. The other point is
that the sign of $q$ gives different effects on waves in propagation
features.

For the positive value of $q$, the oscillation frequency $\omega$
of phase waves gets slightly greater, while that of trigger waves
becomes smaller. Hence, phase waves travel a longer way for the
greater value of $q$ even if they have the same frequency.
Such a case of positive $q$ corresponds to the phenomena of phase
waves in the plasmodium, when we regard the reaction kinetics of
eq.~\ref{eqn: tmts} as the oscillation of some metabolic element like
Ca$^{2+}$. The flow make the plasmodium communicate local information
to the wide area in the cell with phase waves.

When the advection constant $q$ is negative, the opposite situation to
the case of positive $q$ arises. Furthermore, we find out the steep
rise of the frequency in Fig.~\ref{fig: k-omega}. Such steepness
denotes the sharp transition from phase waves to trigger waves with
increasing dimensions of the propagating media.

%---------------------------------------------------------------------
\section{Phase dynamics}
\label{sec: Phase dynamics}

In this section, we show that the effect of the flow comes out through
the coefficient of the nonlinear term in phase dynamics for
reaction-diffusion-advection systems. By means of limit cycle
perturbations, the dynamics of phase waves in ordinary
reaction-diffusion systems for oscillatory media are described by the
Burgers equation\cite{Ortoleva,Kuramoto}. We adopt the similar method
to oscillatory reaction-diffusion equations with advection terms
introduced in section \ref{sec: Model equations}.

We assume that the limit cycle of eq.~\ref{eqn: rda} has the frequency
$\omega_0$. Then a solution of homogeneous oscillation is
\begin{eqnarray*}
  &&{\bf u} = {\bf u}_0(\tau),\quad
  \tau = \omega_0 t,\quad \mbox{where}
  \\
  &&\omega_0 {\bf u}_0' = {\bf f}({\bf u}_0),\quad \mbox{and} \quad
  {\bf u}_0(\tau + 2 \pi) = {\bf u}_0(\tau).
\end{eqnarray*}
Since the system \ref{eqn: rda} is invariant under the time
translation, ${\bf u}_0(\tau + \psi)$ ($\psi$ is an arbitrary
constant) is also a solution of eq.~\ref{eqn: rda}.

Let us consider the dynamics of phase waves in eq.~\ref{eqn: rda} when 
the advection and diffusion terms are small. We introduce multiple
scales and asymptotic expansions,
\begin{eqnarray}
  &&\vec{R} = \sqrt{\epsilon} \vec{r},\quad
  \tau = \omega_0 t,\quad
  T = \epsilon t,
  \nonumber\\
  &&{\bf u} = {\bf u}_0(\tau + \psi) + \epsilon {\bf u} + \cdots,
  \label{eqn: asymptotic expansions}
\end{eqnarray}
where $\epsilon$ is a small parameter and $\psi = \psi(\vec{R}, T)$.
Substitution of eq.~\ref{eqn: asymptotic expansions}
into eq.~\ref{eqn: rda} yields perturbation equations for each order
in $\epsilon$:
\begin{eqnarray}
  &&\omega_0 \frac{\partial {\bf u}_0}{\partial \tau}
  = {\bf f}({\bf u}_0),
  \nonumber\\
  &&L {\bf u}_j = {\bf b}_j,\quad
  L = \omega_0 \frac{\partial}{\partial \tau}
  - \frac{\partial {\bf f}}{\partial {\bf u}}({\bf u}_0),
  \label{eqn: perturbation equations}
\end{eqnarray}
here $j = 1, 2, \ldots$, and ${\bf b}_j$ denotes the inhomogeneous
term of the $j$th order equation. For the first order equation in
eq.~\ref{eqn: perturbation equations}, the inhomogeneous term is
\begin{eqnarray*}
  {\bf b}_1
  &=& - \frac{\partial {\bf u}_0}{\partial T}
  - M \vec{\nabla}_R {\bf u}_0 \cdot \vec{\nabla}_R {\bf u}_0
  + D \vec{\nabla}_R^2 {\bf u}_0
  \\
  &=& - {\bf u}_0' \frac{\partial \psi}{\partial T}
  - M {\bf u}_0' {\bf u}_0' | \vec{\nabla}_R \psi |^2
  \\
  &&  + D {\bf u}_0'' | \vec{\nabla}_R \psi |^2
  + D {\bf u}_0' \vec{\nabla}_R^2 \psi,
\end{eqnarray*}
where $\vec{\nabla}_R$ is the nabla operator in respect to scaled
coordinates $\vec{R}$. Thus the solvability condition for ${\bf u}_1$
gives the dynamics of phase waves:
\begin{equation}
  \frac{\partial \psi}{\partial T}
  = c_1 \vec{\nabla}_R^2 \psi + c_2 | \vec{\nabla}_R \psi |^2.
  \label{eqn: phase dynamics}
\end{equation}
If we use a new dependent variable
$\vec{V} \equiv \vec{\nabla}_R \psi$,
then it satisfies the Burgers equation. The coefficients $c_1$ and
$c_2$ are obtained from the relations,
\begin{eqnarray*}
  &&c_j = \langle {\bf v}^\dagger, {\bf v}_j \rangle
  / \langle {\bf v}^\dagger, {\bf u}_0 \rangle,
  \\
  &&{\bf v}_1 = D {\bf u}_0',\quad
  {\bf v}_2 = D {\bf u}_0'' - M {\bf u}_0' {\bf u}_0',
\end{eqnarray*}
here $\langle {\bf v}^\dagger, {\bf v} \rangle \equiv
\displaystyle\int_0^{2\pi} \!\!{\rm d}\psi ({\bf v}^\dagger, {\bf v})$
and ${\bf v}^\dagger$ is the nontrivial periodic solution to the
adjoint differential equation of $L {\bf v} = {\bf 0}$.
Equation~\ref{eqn: phase dynamics} describes slow and slight
modulation of the homogeneous oscillation with the frequency
$\omega_0$ by the phase $\psi$. We note that the coefficient of the
nonlinear term, $c_2$, show competition between diffusion and
advection.

When we use the quantity $\phi = \omega_0 t + \psi$,
eq.~\ref{eqn: phase dynamics} becomes
\begin{equation}
  \frac{\partial \phi}{\partial t}
  = \omega_0 + c_1 \vec{\nabla}^2 \phi
  + c_2 | \vec{\nabla} \phi |^2.
  \label{eqn: phase}
\end{equation}
The dispersion relation is thus estimated from the phase equation
\ref{eqn: phase} through the wave characteristics
$\omega = \partial \phi / \partial t$ and
$\vec{k} = \vec{\nabla} \phi$ as \cite{Aliev94}
\begin{equation}
  \omega = \omega_0 + c_2 k^2 + \cdots,\quad
  k = | \vec{k} |.
  \label{eqn: dispersion relation}
\end{equation}
Since the scaling of coordinates in the perturbation expansions
\ref{eqn: asymptotic expansions} means spatially slight modulation,
$k = O(\sqrt{\epsilon})$, eq.~\ref{eqn: dispersion relation} is the
Taylor expansion for the dispersion curve, $\omega = \omega(k)$ in the 
vicinity of $k = 0$. Thus, the coefficient of nonlinear term in
eq.~\ref{eqn: phase} is $c_2 = \omega''(0) / 2$.
Here eq.~\ref{eqn: dispersion relation} has no linear term in $k$
because of the reflectional symmetry in space of eq.~\ref{eqn: rda}.
As mentioned above, $c_2$ depends on advection constants as well as
diffusion constants, and hence $c_2$ can take the value of a wide
range.

We point out that the dispersion relation
\ref{eqn: dispersion relation} is only applicable to some of periodic
waves with stationary traveling. It is not adopted to waves with
non-uniform phase gradients \cite{Polezhaev95}.
In such a case, we need to use the phase equation \ref{eqn: phase}, or
analyze eq.~\ref{eqn: rda} directly.

%---------------------------------------------------------------------
\section{Discussion}
\label{sec: Discussion}

By means of numerical calculations for eq.~\ref{eqn: 2rda}, it has
been shown that the flow make remarkable differences in the dispersion
curves or propagation of phase waves. To elucidate such differences
induced by the flow, we consider the spatial scale. The two-component
system \ref{eqn: 2rda} has the diffusion term only in the free
element. Thus, without advection term, the variation of the diffusion
constant is canceled by the scaling of spatial coordinates.
In contrast, such a cancellation is impossible with the advection
term because the advection coefficient is independent of the
diffusion constant.

The deformation of the dispersion curves with the flow term has been
shown for the branch of phase waves by the phase equation derived from
eq.~\ref{eqn: rda}. Furthermore, we have found out another phenomenon
of phase wave mentioned below.

Advection coefficients govern the coefficient $c_2$ of the quadratic
term in the wave number $k$ to
the dispersion relation \ref{eqn: dispersion relation}.
The coefficient $c_2$ is given by the addition of two parts which
stem from the diffusion and advection terms, respectively.
Thus, the competition between the diffusion and advection constants
governs the value of $c_2$. The variation of $c_2$ gives a quadratic
change in the frequency of long waves. This result elucidates the
branch of phase waves ($k \sim 0$) for the dispersion curves obtained
numerically in section~\ref{sec: Dispersion relation}.

Varying the advection constants, we can take $c_2$ negative. Such a
situation seems to be impossible for simple reaction-diffusion
equations without the advection terms. The dispersion curve is convex
for $c_2 < 0$, and hence $\omega = 0$ at some wave number. This means
that the phase wave is frozen and stationary. It sounds strange, and
we need to study these waves in detail. One of such a system is the
$\lambda$-$\omega$ system with advection terms of phase gradients%.
\cite{Yamada00}.

When the coefficient $c_2$ becomes sufficiently small, it may be
possible to derive other types of phase equations by introducing
different scaling of independent variables from ones used in the
present analysis. Furthermore we need to study waves with relaxation
oscillations to know the characteristics of trigger waves with the
flow, because the phase equation is satisfied only in the region of
the phase-wave branch.

%---------------------------------------------------------------------
\section*{Acknowledgments}

This study was supported by The Sumitomo Foundation (Grant
No. 970628), and The Institute of Physical and Chemical Research
(RIKEN) (T. N.).

%---------------------------------------------------------------------

\newpage
%\end{document}
%---------------------------------------------------------------------
% Figure captions

\newpage

\newpage

%%%%%%%%%%%%%%%%%%%%%%%%%%%%%%%%%%%%%%%%%%%%%%%%%%%%%%%%%%%%%%%%%%%%%%
\end{document}